\documentclass[a4paper]{article}
\usepackage{graphicx}
\usepackage{amsmath,amssymb}
\usepackage{indentfirst}%para dar espaço no primeiro parágrafo

\newcommand{\Tr}{{\rm Tr}}

\newcommand{\be}{\begin{equation}}
\newcommand{\ee}{\end{equation}}

\begin{document}

\title{Semiclassical calculation of time delay statistics in chaotic quantum scattering}
\author{Marcel Novaes \\ Instituto de F\'isica, Universidade Federal de Uberl\^andia\\ Uberl\^andia, MG, 38408-100, Brazil}
\date{}

\maketitle
\begin{abstract}

We present a semiclassical calculation, based on classical action correlations, of all moments of the Wigner–Smith time delay matrix, $Q$, in the context of quantum scattering through systems with chaotic dynamics. Our results are valid for broken time reversal symmetry and depend only on the classical dwell time and the number of open channels, $M$, which is arbitrary. Agreement with corresponding random matrix theory reduces to an identity involving some combinatorial concepts, which can be proved in special cases.
\end{abstract}

\section{Introduction}

Wave scattering at energy $E$ can be described by a unitary $S(E)$ matrix, which connects incoming to outgoing amplitudes. We consider a finite region with fully chaotic classical dynamics, connected to the outside world by means of an infinite lead with $M$ open channels, so that $S$ is $M$ dimensional. The dynamics is characterized by a single timescale, the dwell time $\tau_D$, the average time spent in the region by particles randomly injected through the lead. More concretely, if the classical phase-space is filled with random initial conditions, the total mass remaining in the region after
time $t$ decays exponentially as $e^{-t/\tau_D}$.

We focus our attention on the Wigner–Smith time delay matrix \cite{time1,time2,time3},
\be Q(E)=-i\hbar S^\dagger \frac{dS}{dE},\ee
where $\hbar$ is Planck's constant (we use the language of quantum mechanics, but the results are valid for the scattering of other kinds of waves). Its real eigenvalues are the proper time delays of the system, which provide the lifetimes of metastable states \cite{time4,time5}. The normalized trace $\tau_W = \frac{1}{M}\Tr(Q)$ is the Wigner time delay, which is also a measure of the density of states of the open system. 

When the classical dynamics is chaotic, matrix elements of $Q$ are widely fluctating functions of energy and it makes sense to introduce a local energy average. The average value of the Wigner time delay, for example, equals the classical dwell time, $\langle\tau_W \rangle=\tau_D$. More refined statistical information about the time duration of wave scattering is encoded in traces of higher powers, $\Tr(Q^n)$, and more generally in products of such traces, e.g. in quantities like 
\be p_\mu(Q)=\prod_{i=1}^{\ell(\mu)}\Tr(Q^{\mu_i}),\ee
defined in terms of an integer partition $\mu=(\mu_1,\ldots,\mu_{\ell(\mu)})$, i.e. a non-decreasing sequence of $\ell(\mu)$ positive integers. We call such quantities the moments of $Q$.

Within the so-called random matrix theory (RMT) approach, a kind of minimum-information statistical treatment, $Q$ is treated as a random matrix \cite{time4,rmt1,rmt2} and it has been shown that its inverse should be considered as distributed in the Laguerre ensemble \cite{rmt3,rmt4}. This leads to several interesting results \cite{results1,results2,results3,results4,results5,results6,results7,results8,results9}, in particular an explicit formula for $\langle p_\mu(Q) \rangle$ when time-reversal symmetry is not present \cite{eu1} (this was recently generalized for all symmetry classes in \cite{eu2}, but the resulting formulas are not quite as explicit). 

In this work we rely on a semiclassical approach involving sums over scattering rays that produce constructive interference. This has been very successful
in treating transport properties \cite{semi1,semi2,semi3} and energy correlations \cite{delay0,delay1,delay2,delay3}, but the version we employ is the one introduced by Kuipers, Savin and Sieber specifically for time delay \cite{efficient} and which is based on trajectories that enter the system but do not leave. It leads to a perturbative diagrammatic formulation for $\langle p_\mu(Q) \rangle$ in which the contribution of a diagram depends only on $M$ in such a way that an infinite series in powers of $1/M$ is produced. The semiclassical and the RMT approaches are expected to be equivalent, but proving this is a central challenge in the field of quantum chaos. 

We avail ourselves of some recent progress in semiclassical theory by formulating it in terms of auxiliary matrix integrals \cite{matrix1,matrix2,matrix3} and then solving such integrals by invoking concepts from representation theory and combinatorics. This method turns out to be very efficient, in that it leads to a formula for $\langle p_\mu(Q) \rangle$ which is an infinite sum of rational functions of $M$ and is in principle able to provide exact results and not only perturbative ones. This can be done when $\mu$ is of the form $\mu=(n,1,\ldots,1)$ (called a hook), and the result is in perfect agreement with RMT. For more general $\mu$ we can verify this agreement up to several orders in $1/M$ and we conjecture it to be valid in general, leading to a conjectured sum that may be of combinatorial interest.

\section{Semiclassical approximation} 

The authors of \cite{efficient} started from a representation for the Wigner-Smith matrix 
\be Q=\hbar V^\dagger \frac{1}{(E-H_{ef})^\dagger}\frac{1}{(E-H_{ef})}V\ee
in terms of a matrix $V$ coupling the chaotic region to the $M$ open chanels in the lead, and a non-Hermitian effective Hamiltonian  $H_{ef}$. Then, the semiclassical approximation was introduced by means of the Gutzwiller expression \cite{gutz} for the elements of the Green function $(E-H_{ef})^{-1}$,
\be G(r,r',E)\approx\frac{1}{i\hbar\sqrt{2\pi i \hbar}}\sum_\gamma \frac{1}{\sqrt{v_\gamma v_\gamma'|(M_\gamma)_{12}|}}\exp\left(\frac{i}{\hbar}S_\gamma-i\frac{\pi}{2}n_\gamma\right),\ee
where the sum is over all trajectories from $r'$ to $r$ with energy $E$, with $S_\gamma$ the action of $\gamma$, $v_\gamma'$ and $v_\gamma$ the initial and final velocities, $n_\gamma$ the number of conjugate points and $M_\gamma$ the stability matrix that describes linearized motion near the trajectory. 

For the average traces $\langle\Tr(Q^n)\rangle$, for example, they obtain the expression
\be \left\langle\sum_{\vec{i}}\int dr_1\cdots dr_n\sum_{\gamma,\gamma'}A_{\gamma}A^*_{\gamma'}e^{i(S_\gamma-S_{\gamma'})/\hbar}\right\rangle,\ee
where $n$ trajectories $\gamma$ and $n$ trajectories $\gamma'$ enter the chaotic region, with $\gamma_i$ going from channel $i_k$ to end point $r_k$ and $\gamma_k'$ going from channel $i_{k+1}$ to end point $r_k$. The total actions of those two sets of trajectories are $S_\gamma=\sum_k S_{\gamma_k}$ and $S_{\gamma'}=\sum_k S_{\gamma_k'}$, while $A_\gamma=\prod_k A_{\gamma_k}$ is a total stability factor (related to the stability matrix). 

Finally, a stationary phase approximation was performed along with some phase space integrations in order to arrive at a diagrammatic perturbative theory for time delay moments. Diagrams consist of initial and final vertices, corresponding to the lead channels and the trajectory endpoints, together with some internal vertices which correspond to the so-called ``encounters'' which have been recognized to be the mechanism responsible for systematic constructive interference between sets of trajectories, i.e. for classical action correlations. Vertices are connected by edges, corresponding to long stretches of chaotic motion along which the $\gamma$ and $\gamma'$ trajectories are indistinguishable. 

The diagrammatic rules, very similar to the transport ones \cite{semi1,semi2,semi3}, are as follows: summation over each incoming channel gives a factor of $M$; each edge contributes a factor of $1/M$; each vertex gives a factor of $(-M)$, unless it contains an end point; it contributes a factor of $1$ if it contains one end point, and a factor of 0 if it contains more than one end point. The main advantage of this approach compared to previous semiclassical treatments of time delay statistics is that it avoids considering trajectories with different energies. 

Using these rules, Kuipers, Savin and Sieber obtained semiclassical approximations to $\langle\Tr(Q^n)\rangle$, for arbitrary $n$, but restricted to leading orders in a $1/M$ expansion. 

Our method consists in encoding these diagrammatic rules in a properly designed matrix integral. This method was first developed in \cite{matrix1}, where it was applied to the calculation of transport moments. It was more recently applied to systems with tunnel barriers \cite{tunnel} and to time delay calculations via energy correlations \cite{eu3}. In this latter work the method allowed finding exact results for $\langle\Tr(Q^n)\rangle$. However, it could not treat the more general case of $\langle p_\mu(Q)\rangle$. 

The integral we propose to encode the more efficient semiclassical approach of \cite{efficient} for the calculation of $\langle p_\mu(Q)\rangle$, with $\sum_i\mu_i=n$, traditionally denoted $\mu\vdash n$, is this:
\be (M\tau_D)^n\sum_{\vec{i}}\frac{1}{\mathcal{Z}}\int e^{-M\Tr(Z^\dagger Z)}e^{-M\sum_{q\ge 2}\frac{1}{q}\Tr(Z^\dagger Z)^q}\prod_{k=1}^n Z^*_{k,i_k}\left(Z\frac{1}{1-Z^\dagger Z}\right)_{k,i_{\pi(k)}},\ee where $Z$ is a complex $N$-dimensional matrix wih a Gaussian distribution given by the first term in the integrand. The corresponding normalization is given by $ \mathcal{Z}=\int e^{-M\Tr(Z^\dagger Z)}$. The rest of the integrand is to be interpreted as follows: 
\begin{itemize}
\item the term $Z^*_{k,i_k}$ represents the trajectory $\gamma_k$ going from channel $i_k$ to end point $r_k$;

\item the term $\left(Z\frac{1}{1-Z^\dagger Z}\right)_{k,i_{\pi(k)}}$ represents the trajectory $\gamma_k'$ going from channel $i_{\pi(k)}$ to end point $r_k$; here $\pi$ is any permutation with cycle type $\mu$. The geometric series produces vertices of any valence that contain an end point;

\item the term $e^{-M\sum_{q\ge 2}\frac{1}{q}\Tr(Z^\dagger Z)^q}$, when Taylor expanded, produces all possible vertices without end points, each accompanied by a factor $(-M)$.

\item the sum over $i_1,\ldots,i_n$ takes into account all possible channels through which a trajectory may enter the chaotic region (remember that in this theory the trajectories do not leave).
\end{itemize}

When the integral is computed using Wick's rule \cite{wick1,wick2,wick3}, each edge will be accompanied by a factor $1/M$, as it should, so the diagrammatical formulation of this integral indeed coincides with the semiclassical rules of \cite{efficient}. 

Except that, when Wick’s rule is applied, some of the resulting diagrams may contain closed cycles. These would correspond to periodic orbits in the semiclassical theory, but such orbits should not be present. The number of such cycles is controlled by the dimension $N$: the contribution of a diagram with $t$ periodic orbits is proportional to $N^t$. Therefore, to exclude them we consider the part of the result that is independent of $N$ or, equivalently, we take the limit $N\to 0$.

\section{Computing the integral}

Introduce the singular value decomposition $Z=UDV$, with $U$ and $V$ in the unitary group $\mathcal{U}(N)$ and $D$ a real and non-negative diagonal matrix. The jacobian of this transformation is $dZ=dUdVdX\Delta(X)^2$ with $X=D^2$ and $\Delta(X)$ being the Vandermonde,
\be \Delta(X)=\prod_{j<k}(x_k-x_j).\ee

Then, the terms $\prod_kZ^*_{k,i_k}\left(Z\frac{1}{1-Z^\dagger Z}\right)_{k,i_{\pi(k)}}$ become
\be\label{chan} \sum_{\vec{a},\vec{c},\vec{i}}\frac{D_{c_k}}{1-x_{c_k}}D_{a_k}\int_{\mathcal{U}(N)} \prod_k U_{k,c_k}U^*_{k,a_k}dU \int_{\mathcal{U}(N)} \prod_k V_{c_k,i_{\pi(k)}}V^*_{a_k,i_k}dV.\ee 
The unitary integrals are computed using Weingarten calculus \cite{weing}. In order to state it, let us introduce some concepts from representation theory (see for example \cite{mac,stan} for detailed accounts). Denote by $S_n$ the permutation group of $n$ symbols and by $\chi_\lambda(\pi)$ the character of its irreducible representation labelled by the integer partition $\lambda$. 
In particular, let $d_\lambda=\chi_\lambda(1)$ be the dimension of the representation. Such characters satisfy the orthogonality relation
\be\label{orth} \frac{1}{n!}\sum_{\beta \in S_n}\chi_\rho(\alpha\beta)\chi_\delta(\pi\beta)=\frac{\chi_\rho(\alpha\pi)}{d_\rho}\delta_{\rho,\delta}.\ee A partition $\lambda$ may be represented by its Young diagram, an arrangement of boxes at positions $(i,j)$ with $1\le i\le \ell(\lambda)$ and $1\le j\le \lambda_i$. The content of box $(i,j)$ is $j-i$ and the content polynomial of $\lambda$ is
\be\label{contp} [x]^\lambda=\prod_{(i,j)\in \lambda}(x+j-i)=\prod_{i=1}^{\ell(\lambda)}\frac{\Gamma(x+\lambda_i-i+1)}{\Gamma(x-i+1)}. \ee

All the above concepts appear in the expression for the unitary integrals, 
\be \int_{\mathcal{U}(N)} \prod_k U_{k,c_k}U^*_{k,a_k}dU=\frac{1}{n!}\sum_{\sigma,\tau\in S_n} \sum_{\lambda\vdash n}\frac{d_\lambda\chi_\lambda(\sigma^{-1}\tau)}{[N]^{\lambda}}\delta_\sigma(\vec{n},\vec{n})\delta_\tau(\vec{c},\vec{a}),\ee where $\vec{n}=(1,\ldots,n)$ and the function $\delta_\tau(\vec{c},\vec{a})$ equals 1 if the list $\vec{a}$ is identical with list $\vec{c}$ permuted by $\tau$, and vanishes otherwise. Since the list $\vec{n}$ does not have any repeated elements, $\sigma$ must be the identity. Likewise, we have
\be \int_{\mathcal{U}(N)} \prod_k V_{c_k,i_{\pi(k)}}V^*_{a_k,i_k}dV=\frac{1}{n!}\sum_{\alpha,\beta\in S_n} \sum_{\rho\vdash n}\frac{d_\rho\chi_\rho(\alpha^{-1}\beta)}{[N]^{\rho}}\delta_\alpha(\vec{c},\vec{a})\delta_\beta(\vec{i},\pi(\vec{i})).\ee

The sum over channels is simply
\be \sum_{\vec{i}}\delta_\beta(\vec{i},\pi(\vec{i}))=M^{\ell(\pi^{-1}\beta)}=p_{\pi^{-1}\beta}(1^M),\ee
where $1^M$ is the $M$-dimensional identity matrix. The function $p_\mu(X)$ is a symmetric polynomial in the eigenvalues of $X$ and, as such, can be written as a linear combination of Schur polynomials. This decomposition is well known to be given as
\be\label{decomp} p_{\beta}(X)=\sum_{\delta\vdash n}\chi_\delta(\beta)s_\delta(X).\ee
On the other hand, the sum over $\vec{a}$ and $\vec{c}$ is
\be \sum_{\vec{a},\vec{c}}\delta_\tau(\vec{c},\vec{a})\delta_\alpha(\vec{c},\vec{a})
\frac{D_{c_k}}{1-x_{c_k}}D_{a_k}=p_{\alpha^{-1}\tau}\left(\frac{X}{1-X}\right).\ee
Using the orthogonality relation (\ref{orth}), the quantity in Eq.(\ref{chan}) becomes
\be \frac{1}{n!}\sum_{\tau,\alpha\in S_n} \sum_{\lambda,\rho\vdash n}\frac{d_\lambda\chi_\lambda(\tau)}{[N]^{\lambda}}\frac{\chi_\rho(\alpha\pi^{-1})}{[N]^{\rho}}s_\rho(1^M)p_{\alpha^{-1}\tau}\left(\frac{X}{1-X}\right).\ee
Using (\ref{decomp}) two more times, this reduces to
\be\sum_\lambda \chi_\lambda(\pi) s_\lambda\left(\frac{X}{1-X}\right)\frac{[M]^\lambda}{([N]^\lambda)^2},\ee
where we have used $s_\lambda(1^M)=\frac{d_\lambda}{n!}[M]^\lambda$ (see \cite{mac,stan}).

Having performed the integrals over the angular degrees of freedom, we are left with the eigenvalue integral
\be \label{matrix}(M\tau_D)^n\lim_{N\to 0}\sum_{\lambda\vdash n}\chi_\lambda(\mu)\frac{[M]^\lambda}{([N]^\lambda)^2}\frac{1}{\mathcal{Z}}\int_0^1 \det(1-X)^M\Delta(X)^2s_\lambda\left(\frac{X}{1-X}\right)dX,\ee
where we have summed over $q$ in the exponent of the integrand, and used that $e^{\Tr\log(A)}=\det(A)$. In order to carry out this integration, we must first express $s_\lambda\left(\frac{X}{1-X}\right)$ as a linear combination of Schur polynomials of $X$. This is given by the ``geometric series'' generalization
\be s_\lambda\left(\frac{X}{1-X}\right)=\sum_{\rho\supset\lambda}C_{\lambda,\rho}s_\rho(X),\ee
where $\rho\supset\lambda$ means that the Young diagram of $\rho$ contains that of $\lambda$ and
\be C_{\lambda,\rho}=\det\left[\binom{\rho_j-j}{\lambda_i-i}\right].\ee 

Then we can use the Selberg integral \cite{selberg}
\be \int_0^1 \det(1-X)^M\Delta(X)^2s_\rho(X)dX=\frac{d_\rho}{|\rho|!}([N]^\rho)^2\prod_{j=1}^N \frac{j!(j-1)!\Gamma(M+N-j+1)}{\Gamma(\rho_j+M+2N-j+1)}.\ee
A different version of Selberg integral can also be used to compute the normalization constant
\be \int_0^\infty e^{-M\Tr(X)}\Delta(X)^2dX=M^{-N^2}\prod_{j=1}^N j!(j-1)!.\ee
Dividing these two results, we see that our matrix integral contains the factor
\be \left(\frac{[N]^\rho}{[N]^\lambda}\right)^2M^{N^2}\prod_{j=1}^N \frac{\Gamma(M+N-j+1)}{\Gamma(\rho_j+M+2N-j+1)}.\ee 
At this point we take the limit $N\to 0$, after which the above quantity becomes simply
\be \left(\frac{t^\rho}{t^\lambda}\right)^2\frac{1}{[M]^\rho},\ee
where $t^\rho$ is the product of all non-zero contents of the partition $\rho$,
\be t^\rho=\prod_{i=1}^{\ell(\rho)}\prod_{j\neq i}(j-i).\ee This is  derived directly from (\ref{contp}). Notice that the number of boxes with zero content must be equal in the Young diagrams of $\rho$ and of $\lambda$. 

Finally, our expression for the time delay moments is
\be \langle p_\mu(Q)\rangle=(M\tau_D)^n\sum_{\lambda\vdash n}\chi_\lambda(\mu)\frac{[M]^\lambda}{t_\lambda^2}\sum_{\rho\supset\lambda} C_{\lambda,\rho}\frac{d_\rho t_\rho^2}{|\rho|![M]^\rho}.\ee Alternatively, we may express the result in terms of average Schur polynomials as
\be\label{final} \langle s_\lambda(Q)\rangle=(M\tau_D)^n\frac{[M]^\lambda}{t_\lambda^2}\sum_{\rho\supset\lambda} C_{\lambda,\rho}\frac{d_\rho t_\rho^2}{|\rho|![M]^\rho}.\ee 

\section{Agreement with random matrix theory}

The result obtained from RMT for time delay statistics is, for a given $\lambda\vdash n$, equal to \cite{eu1,eu2}
\be \langle s_\lambda(Q)\rangle_{RMT}=(M\tau_D)^n\frac{d_\lambda [M]^\lambda}{n![M]_\lambda},\ee where the denominator is given by
\be [M]_\lambda=\prod_{i=1}^{\ell(\lambda)}\frac{\Gamma(M+i+1)}{\Gamma(M-\lambda_i+i+1)}. \ee

Therefore, our semiclassical result (\ref{final}) coincides with the RMT prediction, providing it with a more solid derivation from first principles, provided that the following identity holds:
\be \sum_{\rho\supset\lambda} C_{\lambda,\rho}\frac{d_\rho t_\rho^2}{|\rho|![M]^\rho}=\frac{d_\lambda t_\lambda^2}{n![M]_\lambda}.\ee
Let $\lambda'$ be the conjugate partition to $\lambda$, obtained by transposing its Young diagram, i.e. turning columns into rows and vice-versa. Then it follows that $[M]_\lambda=[M]^{\lambda'}$, and we can also express our identity in terms of Schur polynomials in the very simple form
\be\label{id} \sum_{\rho\supset\lambda} C_{\lambda,\rho}\frac{t_\rho^2}{s_\rho(1^M)}=\frac{t_\lambda^2}{s_{\lambda'}(1^M)}.\ee

However, simplicity aside, this identity does not seem easy to prove. Notice that the right-hand side has poles for all $M<\lambda_1$, so we must assume that $M\ge \lambda_1$. The left-hand side has, at first sight, poles at all positive integer values of $M$, so inserting a concrete value of $M$ only makes sense after the summation has been performed and all spurious poles have been cancelled. 

Another way to look at it is that both sides have the same $1/M$ expansion. 

The simplest case, for example, is $\lambda=(1)$. The right-hand side is then given by $1/M$. For the left hand side, let us take $\rho\vdash r$ with $\rho_1=n+1$ and $\rho_i=1$ for $1<i\le r-n$, with $0\le n< r$. Then if we sum over $r$ up to $2$ we have
\be \sum_{r=1}^2\sum_{n=0}^{r-1} C_{\lambda,\rho}\frac{t_\rho^2}{s_\rho(1^M)}=-\frac{1}{(M-1)M(M+1)}=\frac{1}{M}+O(M^{-3}),\ee if we go up to 3 we have
\be \sum_{r=1}^3\sum_{n=0}^{r-1} C_{\lambda,\rho}\frac{t_\rho^2}{s_\rho(1^M)}=\frac{8}{(M-2)(M-1)M(M+1)(M+2)}=\frac{1}{M}+O(M^{-5}),\ee
and so on.

We see that the partial sums for the left hand side indeed have spurious poles, but asymptotically they approach the right hand side. 

As a matter of fact, the required identity can indeed be proved when $\lambda$ is a hook, i.e. a partition of the form $(A,1^a)$, as we show next. Unfortunately, we must leave the validity of Eq.(\ref{id}) in the general case as a conjecture (we have checked it extensively).

\subsection{Moments of hook shape}

If $\lambda=(A,1^a)$ is a hook, so must be $\rho=(B,1^b)$, because, as we hve seen, these two partitions must have the same number of zero contents. In this case it is easy to see that the product of non-zero contents is $t^\lambda=(-1)^a a!(A-1)!$, and the dimension of the corresponding irreducible representation is $d_\lambda=\binom{A+a-1}{a}$. The content polynomial is the rising factorial $(M-a)^{(A+a)}$. Moreover, the coefficient $C_{\lambda\rho}$ is also known,
\be C_{(A,1^a),(B,1^b)}=(-1)^{a+b}{b\choose a} {B-1\choose A-1}.\ee 

Thefore, our identity becomes very explicit:
\be\label{exp} \sum_{B\ge A,b\ge a} (-1)^{a+b}{b\choose a} {B-1\choose A-1}\frac{(B-1)!b!}{(B+b)(M-b)^{(B+b)}}=\frac{(A-1)!a!}{(A+a)(M-A+1)^{(A+a)}}.\ee 

This can be proved as follows \cite{mo}. Using the Euler beta integral, 

\begin{align}
&\sum_{B, b} (-1)^{a+b}{b\choose a} {B-1\choose A-1}\frac{1}{(B+b) (M+B-1)  }\binom{B+b-1}{b}^{-1}\binom{M+B-2}{B+b-1}^{-1}\\
&= \sum_{B, b} (-1)^{a+b}{b\choose a} {B-1\choose A-1} \int_0^1 \int_0^1 (1-y)^{M-b-1} y^{B+b-1} (1-z)^bz^{B-1}\,{\rm d}y\,{\rm d}z\\
&= -\int_0^1 \int_0^1 (1-y)^{M} y^{-2}(z(1-z))^{-1} \left( \frac{y(1-z)}{1-yz} \right)^{a+1} \left(\frac{yz}{1-yz}\right)^{A}\,{\rm d}y\,{\rm d}z\\
&= -\int_0^1 \int_0^1 (1-y)^{M} (yz)^{A-1} (y-yz)^a (1-yz)^{-(A+a+1)}\,{\rm d}y\,{\rm d}z.
\end{align}

Substituting $(u,v)=(\frac{1-y}{1-yz},yz)$, or $(y,z) = (1-u(1-v),\frac{v}{1-u(1-y)})$ we further get 
\begin{align}
& \int_0^1 \int_0^1 u^M (1-u)^a v^{A-1} (1-v)^{M-A} \frac1{1-u(1-v)}\,{\rm d}u\,{\rm d}v\\
&= \sum_{k\geq 0}\int_0^1 \int_0^1 u^{M+k} (1-u)^a v^{A-1} (1-v)^{M-A+k}\,{\rm d}u\,{\rm d}v \\
&= \sum_{k\geq 0} \frac{1}{(a+1)A}\binom{M+k+a+1}{a+1}^{-1}\binom{M+k}{A}^{-1}  \\
&= \sum_{k\geq 0} \frac{1}{(a+1)A}\binom{A+a+1}{A}^{-1} \binom{M+k+a+1}{A+a+1}^{-1},\\
&= \frac{M+a+1}{(a+1)A (A+a) }\binom{A+a+1}{A}^{-1}\binom{M+a+1}{A+a+1}^{-1} \\
&= \frac{1}{(A+a)^2}\binom{A+a-1}{a}^{-1}\binom{M+a}{A+a}^{-1}=\frac{(A-1)!a!}{(A+a)(M-A+1)^{(A+a)}}.
\end{align}

\section*{Acknowledgments}

Financial support from FAPEMIG, Grant PPM-00126-17, is gratefully acknowledged. I thank Max Alekseyev for providing the proof of equation (\ref{exp}).

\end{document}